\documentclass[twocolumn]{aastex63}
\usepackage{graphicx}
\usepackage{color}
\usepackage[space]{grffile}
\usepackage{latexsym}
\usepackage{textcomp}
\usepackage{longtable}
\usepackage{multirow,booktabs}
\usepackage{amsfonts,amsmath,amssymb}
\usepackage{url}
\usepackage{hyperref}
\usepackage[english]{babel}
\usepackage{graphicx}
\usepackage[space]{grffile}
\usepackage{url}
\usepackage[utf8]{inputenc}
\usepackage{hyperref}
\usepackage{longtable}
\usepackage{natbib}
\usepackage{bm}
\usepackage{xcolor}

\begin{document}

\shorttitle{Scatter in SMHM}
\shortauthors{Munshi et al.}

\title[Quantifying scatter in galaxy formation at the lowest masses]{Quantifying scatter in galaxy formation at the lowest masses}

\correspondingauthor{Ferah Munshi}
\email{fdm@ou.edu}

\author{Ferah Munshi}
\affiliation{Homer L. Dodge Department of Physics \& Astronomy, University of Oklahoma, 440 W. Brooks St., Norman, OK 73019, USA}
\affiliation{Department of Physics \& Astronomy, Rutgers, The State University of New Jersey, 136 Frelinghuysen Road, Piscataway, NJ 08854, USA}
\affiliation{Department of Physics and Astronomy, Vanderbilt University, PMB 401807, Nashville, TN 37206}

\author{Alyson M. Brooks}
\affiliation{Department of Physics \& Astronomy, Rutgers, The State University of New Jersey, 136 Frelinghuysen Road, Piscataway, NJ 08854, USA}

\author{Elaad Applebaum}
\affiliation{Department of Physics \& Astronomy, Rutgers, The State University of New Jersey, 136 Frelinghuysen Road, Piscataway, NJ 08854, USA}

\author{Charlotte R. Christensen}
\affiliation{Department of Physics \& Astronomy,  Grinnell College Grinnell, IA 50112}

\author{Jordan P. Sligh}
\affiliation{Homer L. Dodge Department of Physics \& Astronomy, University of Oklahoma, 440 W. Brooks St., Norman, OK 73019, USA}
 
\author{T. Quinn}
\affiliation{Astronomy Department, University of Washington, Box 351580, Seattle, WA, 98195-1580}


\label{firstpage}

\begin{abstract}

We predict the stellar mass -- halo mass (SMHM) relationship for dwarf galaxies, using simulated galaxies with peak halo masses of M$_{\rm peak} = 10^{11}$ M$_{\odot}$ down into the ultra-faint dwarf range to M$_{\rm peak} =$ 10$^7$ M$_{\odot}$.  Our simulated dwarfs have stellar masses of M$_{\rm star} = $ 790 M$_{\odot}$ to $8.2 \times 10^8$ M$_{\odot}$, with corresponding $V$-band magnitudes from $-2$ to $-18.5$. For M$_{\rm peak} > 10^{10}$ M$_{\odot}$, the simulated SMHM relationship agrees with literature determinations, including exhibiting a small scatter of 0.3 dex.   However, the scatter in the SMHM relation increases for lower-mass halos.  We first present results for well-resolved halos that contain a simulated stellar population,
but recognize that whether a halo hosts a galaxy is inherently mass resolution dependent.  We thus adopt a probabilistic model to populate ``dark’’ halos below our resolution limit to predict an ``intrinsic’’ slope and scatter for the SMHM relation.  We fit linearly growing log-normal scatter in stellar mass, which grows to more than 1 dex at M$_{\rm peak}$ $=$ 10$^8$ M$_{\odot}$. 
At the faintest end of the SMHM relation probed by our simulations, a galaxy cannot be assigned a unique halo mass based solely on its luminosity.  Instead, we provide a formula to stochastically populate low-mass halos following our results.  Finally, we show that our growing log-normal scatter steepens the faint-end slope of the predicted stellar mass function.

\end{abstract}%

\section{Introduction}

In the $\Lambda$ Cold Dark Matter ($\Lambda$CDM) paradigm of cosmological structure formation, dwarf galaxies are predicted to be the smallest, most abundant, yet least luminous galaxies in the Universe. 
Attempts to link dwarf galaxies to their parent dark matter halos via abundance matching have led to discrepancies between theory and observations \citep[e.g.,][]{Ferrero2012,GK2014b, Papastergis2015, BDC2015}.  
Abundance matching matches a stellar mass or luminosity at a given abundance to dark matter halos with the same abundance,  derived from a dark matter-only simulation.  A monotonic relationship is generally assumed \citep{guo10,Behroozi2013,Moster2013}.  Critically, abundance matching also assumes that every dark matter halo is occupied by a galaxy.  Abundance matching studies generally yield fair agreement for the stellar mass-to-halo mass (SMHM) relation for halos of masses $\gtrsim 10^{11}$ M$_{\odot}$.  Additionally, for halos of roughly Milky Way mass and greater, abundance matching also reproduces clustering statistics \citep[e.g.,][]{ConroyWechsler2009, Wechsler2018}.  

However, derivations of the SMHM relation at lower masses have yielded discrepancies \citep[e.g.,][]{Moster2013, Behroozi2013, Brook2014, GK2014, Read2017, Jethwa2018}. If the SMHM relation has the form M$_{\rm star} \propto$ M$^{\alpha}_{\rm halo}$, the range of derived $\alpha$ varies from 1.4 to 3.1 for galaxies smaller than M$_{\rm halo} < 10^{11.5}$ M$_{\odot}$.
However, \citet{Moster2013} and \citet{Behroozi2013} did not have data in order to derive the SMHM relation below stellar masses of a few $\times$10$^{7}$ M$_{\odot}$.  While their results are unconstrained at
lower masses, the slopes at their lowest measured  mass were quite different, $\alpha = 1.4$ \citep{Behroozi2013} versus $\alpha = 2.4$ \citep{Moster2013}.  \citet{Read2017} find a fairly shallow relation between $10^{7}$ $<$ M$_{\rm star}$/M$_{\odot}$ $<$ $10^{9}$, with $\alpha \approx 1.2$, for field dwarfs in SDSS.  They also derive the halo masses of isolated, local dwarfs via rotation curve fitting \citep{Read2016b} and find that the SMHM relation for the individual galaxies is well described by their derived SMHM relation from abundance matching, including when extrapolated to lower stellar masses.  They attribute this match to the fact that they use an isolated galaxy sample, arguing that including galaxies processed in a group environment leads to a steeper SMHM relation.  \citet{Brook2014} and \citet{GK2014}, on the other hand, used Local Group galaxy data to determine the SMHM relation at 10$^6 < $M$_{\rm star}$/M$_{\odot} < 10^8$.  Again, they came to quite different conclusions about the value of $\alpha$, 3.1 in \citet{Brook2014} and 1.9 in \citet{GK2014}.\footnote{Though the slope is dependent on the normalization at higher masses, and the values come into better agreement when a consistent normalization is adopted, see \citet{GK2016}.}   

Similarly, both \citet{Jethwa2018} and \citet{Nadler2020} examine the SMHM using Local Group data, specifically using Milky Way satellites.  However, unlike \citet{Brook2014} and \citet{GK2014}, they do not require that every dark matter halo contains a galaxy. \citet{Jethwa2018} explore a range of assumptions, from the standard abundance matching, to models that allow unoccupied halos and varying forms of the stellar mass function.  \citet{Nadler2020} use a more constrained model, building on earlier work to determine which subhalos are likely to contain luminous satellites, and the known selection function of current surveys to determine which satellites should be detected.  Both \citet{Jethwa2018} and \citet{Nadler2020} found that most halos must be occupied down to M$_{\rm peak} \sim 10^8$ M$_{\odot}$.  It is common in abundance matching to assume that all halos continue to host galaxies.

The idea that all halos down to M$_{\rm peak} \sim 10^8$ M$_{\odot}$ are occupied is contrary to traditional wisdom, which holds that reionization may stifle star formation entirely in such low mass halos \citep[e.g.,][]{gnedin00, g07, Okamoto2008, Brooks2013}. 
Simulators generally find that reionization causes low mass halos to be entirely devoid of stars, or contain so few as to be beyond detection \citep[e.g.,][]{OShea2015, Ocvirk2016, Ocvirk2020, Wheeler2019, Katz2020}. 
However, the mass scale at which this occurs remains an open question.  \citet{Sawala2016} used simulations representative of the Local Group to show that the fraction of halos which are not populated (not luminous) increases drastically for $z=0$ halo masses less than $10^9$ M$_{\odot}$, suggesting a plummeting galaxy formation efficiency at these mass scales. \citet{Fitts2018} find a similar drop in galaxy formation below M$_{\rm halo} = $ $10^9$ M$_{\odot}$.  \citet{Ocvirk2020} also find a drop below this halo mass, though at $z=6$.  However, other simulators have found that galaxies populate lower mass halos, closer to M$_{\rm halo} \sim $ $10^8$ M$_{\odot}$ \citep{Wheeler2019, Revaz2018, Katz2020}.  Simulations of the low mass halos at high $z$ have shown that stars can form in even lower mass halos at $z > 6$ \citep[e.g.,][]{Wise2012, Cote2018, Latif2019, Skinner2020, Nakatani2020}. 

In addition to the slope of the SMHM relation and the occupation fraction of low mass halos, there is also uncertainty in the (intrinsic) scatter in the SMHM  relation. The scatter at the high mass end of the SMHM relation is consistently measured to be relatively small, $\sim$0.2 dex  \citep{Behroozi2013, Reddick2013, Eagle2017, Kravtsov2018}, though see \citet{Taylor2020}.  However, the scatter at the low mass end may be much larger.  Observations indicate that dwarf galaxies over a range of stellar masses may all occupy dark matter halos with a narrow range of masses \citep[e.g., see Figure 1 of][]{Klypin2015, Ferrero2012, strigari08}.  \citet{Fitts2017} simulated 15 galaxies all with roughly the same halo mass at $z=0$ ($\approx$ 10$^{10}$ M$_{\odot}$), yet the stellar content varied by two orders of magnitude (10$^{5-7}$ M$_{\odot}$).  The star formation history at dwarf galaxy scales is likely to depend on the mass accretion history of the halo \citep{BZ2014, Weisz2015a, Sawala2016}.  \citet{Rey2019} showed that the stellar content of a {\it single} dark matter halo (with the same final halo mass at $z=0$) could vary based on its growth rate.  If the halo grew rapidly and was more massive at early times, it contained a higher stellar mass at $z=0$, by as much as an order of magnitude.  The results for the single halo in \citet{Rey2019} represent a lower limit on how much scatter in stellar mass might be expected at a fixed halo mass.

\citet{GK2016} explored the implications of scatter within M$_{\rm star}$ at a given M$_{\rm halo}$ using Local Group data down to M$_{\rm star} \sim 10^5$ M$_{\odot}$.  They demonstrated that there is a degeneracy between the slope and the scatter of the SMHM relation when using the SMHM relation to derive the stellar mass function.  Small halos are more likely to scatter to large stellar masses due to the rapidly rising mass function in a $\Lambda$CDM cosmology.  Hence, large scatter requires a steeper SMHM slope in order to reproduce the observed stellar mass function. A steeper stellar mass function may, e.g., alleviate the mismatch between the number of predicted classical dwarfs in CDM with what is observed \citep{GK2016}. On the other hand, \citet{Nadler2020} find that the scatter must be small in order to not overpopulate the observed satellite luminosity function (LF) of the Milky Way. The inherent steepness of the LF at dwarf scales also has implications for how reionization proceeded.  In order for galaxies to reionize the Universe, the UVLF likely has to maintain a steep slope down to UV magnitudes as faint as about $-10$ \citep[e.g.,][]{Kuhlen2012, Robertson2013, Weisz2017}.
Measurements of lensed dwarfs at $1 < z < 3$~show no break in the UVLF down to a UV magnitude of roughly $-14$ \citep{Alavi2016}.  Results from lower resolution simulations using the code adopted in this work 
show that a LF with no break down to $-15$ can reionize the Universe at a rate consistent with observations \citep{Anderson2017}.  In this work we quantify the slope and scatter of the SMHM relation inherent in our simulations down to the ultra-faint dwarf range, and explore the impact on the slope of the stellar mass function.

A number of simulators have used baryonic simulations to show that they match the derived SMHM relation at halo masses above 10$^{10}$ M$_{\odot}$ \citep[e.g.,][]{Brook2012, Aumer2013, FIRE,Munshi13}, but only a few have examined the SMHM  of simulated dwarf galaxies below this mass \citep{Munshi13, Shen2014, Onorbe2015, Sawala2015}.  
Most works do not examine a large enough sample of dwarfs to be able to define a SMHM relation, but can instead only compare to extrapolated abundance matching results. \citet{Sawala2015} studied a larger population of dwarfs, tracing halos down to $\sim10^8$ M$_{\odot}$.  They showed that an abundance matching that used only {\it populated} halos leads to placing higher stellar mass galaxies into halos than traditional abundance matching \citep[see also][]{Jethwa2018}. 
Considering only occupied halos and the fact that baryons alter the halo mass function \citep{Sawala2013, Munshi13, Benson2020} 
leads to a SMHM relation that indicates the typical halo mass at a given stellar mass, unlike traditional abundance matching that indicates the typical stellar mass at a given halo mass. Thus, taking into account baryonic effects leads to a SMHM relation that is much shallower at the low mass end than found by traditional abundance matching.

In this paper we address the question: how do galaxies populate low mass dark matter halos?
We do this using a suite of zoomed-in simulations that contain over 200 dwarfs, the Marvel-ous Dwarfs and Justice League simulations. These simulations were purposely designed to yield a large sample of dwarf galaxies, from LMC-mass at the most massive end, down into the ultra-faint dwarf (UFD) galaxy range for the first time at the low mass end. We examine both field dwarfs and their satellites, and satellites around Milky Way-mass galaxies. However, we make a radical departure from earlier works that assumed that their simulated ``dark'' halos were those impacted by reionization.  Many simulators treat these dark halos as a prediction of the simulation.  Instead, we assume that our dark halos are impacted by our resolution limit, and explore the instrinsic slope and scatter of the SMHM relation including these unresolved galaxies.

We present our new suites of high resolution simulations in Section \ref{Sims}. In Section \ref{Results} we show that the scatter in the SMHM relation grows as halo mass decreases.  We quantify the slope and scatter of the SMHM relation, first for only the galaxies that are well-resolved in our simulation, and then extrapolated to include unresolved galaxies.  We demonstrate the impact of our predicted scatter on the resulting stellar mass function that can be tested by, e.g., the Vera Rubin Observatory's Legacy Survey of Space \& Time (LSST).  In Section \ref{Discuss} we discuss other factors that we have not necessarily explored in this paper but may also impact the resulting SMHM relation.
We summarize our results in Section \ref{Summary}. 

\section{Simulations}\label{Sims}

For the first time, we present the full sample of 211 dwarf galaxies including both the ``Marvel-ous Dwarfs'' zoom simulations, along with the ``DC Justice League'' Milky Way-mass zoom simulations.  The ``Marvel-ous Dwarfs'' (hereafter Marvel) are slightly different than traditional zoom simulations, which generally select one halo of interest and place the highest resolution particles on that halo out to a few virial radii.  For the Marvel runs, we instead selected regions of the Universe that contain dozens of dwarf galaxies and ran the entire region as a zoom-in simulation, with the goal of generating one of the largest samples of simulated dwarf galaxies at incredibly high resolution.  We ran four such simulations (named CptMarvel, Elektra, Rogue, and Storm) within a WMAP3 cosmology \citep{spergel07}, each containing a dozen to a few dozen dwarfs.  The most massive galaxies in the Marvel runs are $\sim$LMC-mass ($\sim$10$^{11}$ M$_{\odot}$ in halo mass), but the high resolution of the simulations (60 pc force resolution, gas, initial star, and dark matter masses of 1410 M$_{\odot}$, 420 M$_{\odot}$, and 6650 M$_{\odot}$, respectively) allows galaxies as low as M$_{\rm star} \sim$3000 M$_{\odot}$ (UFDs) to be resolved. A total of 68 Marvel dwarfs are used in this work, 11 of which are satellites (within the virial radius) of dwarf galaxies. 

The regions selected for the Marvel runs are roughly 1.5 to 7 Mpc away from a Milky Way-mass galaxy and can be considered representative of the Local Volume.  To complement these regions, we also use four zoom simulations of Milky Way-mass galaxies and their surrounding environments (out to $\sim$1 Mpc), the DC Justice League simulations (named after the first four women who have served on the US Supreme Court; Sandra, Ruth, Sonia, and Elena). 
Two of the DC Justice League simulations (Ruth and Sonia) are run at ``Near Mint'' (NM) resolution, which is slightly lower resolution than the Marvel dwarfs (170 pc force resolution, initial gas, initial star, and dark matter masses of 2.7$\times$10$^4$ M$_{\odot}$, 8000 M$_{\odot}$, and 4.2$\times$10$^4$ M$_{\odot}$, respectively) within a Planck cosmology \citep{Planck2016}. However, we include dwarfs from the ``Mint'' resolution DC Justice League (Sandra and Elena) presented in \citet{Applebaum2021} which have a resolution within a factor of two of Marvel (87 pc force resolution and gas, dark, and initial star particle masses of \mbox{3310, 17900, and 994 M$_\sun$}, respectively). This combined ``Mint'' and ``Near-Mint'' set of 4 simulations yields 143 dwarfs: 64 field dwarfs \citep[47 of which are backsplash dwarfs of the Milky Way-mass hosts as defined in][]{Applebaum2021} and 79 satellites.
\footnote{Due to different criteria for inclusion and different halo definitions, our sample, while overlapping with \citet{Applebaum2021}, includes fewer galaxies from Sandra and Elena.} 

Both suites of simulations were run with the N-Body + SPH code {\sc ChaNGa} \citep{Menon2015}. \textsc{ChaNGa} adopts the hydrodynamic modules of \textsc{Gasoline2} \citep{wadsley04, Wadsley2017} but uses a faster gravity solver, as well as the \textsc{charm++} \citep{Kale1993} runtime system for dynamic load balancing and communication.  This allows \textsc{ChaNGa} to scale up to thousands of cores.  It is the excellent scalability of \textsc{ChaNGa} that allowed the Marvel simulation suite and the ``Mint'' DC Justice League simulations to be run.

Both sets of simulations utilize the gas cooling and star formation scheme introduced in \citet{Christensen2012}.  Metal line cooling and the diffusion of metals is included \citep{shen10}, and the non-equilibrium formation and destruction of molecular hydrogen, H$_2$, is followed.  We apply a uniform, time-dependent UV field from \cite{HM2012} in order to model photoionization and heating, and the Lyman-Werner radiation from young stars is tracked.  Star formation is restricted to occur only in the presence of H$_2$.

Star formation occurs stochastically when gas particles become cold ($T < 1000$ K) and dense ($n > 0.1$  $m_H$  cm$^{-3}$).  Although the density threshold is low, in practice the requirement that H$_2$ be present restricts stars to forming only in gas that reaches a density threshold $n > 100$ $m_H$ $\rm cm^{-3}$.    
The probability, $p$, of spawning a star particle in a time $\Delta t$ is a function of the local dynamical time $t_{form}$:
\begin{equation}
p = \frac{m_{gas}}{m_{\rm star}} \left (1 - e^{-c{_0^*} X_{H_2} \Delta t/t_{form}} \right)
\label{eqn1}
\end{equation}
where $m_{gas}$ is the mass of the gas particle and $m_{\rm star}$ is the initial mass of the potential star particle. A star formation efficiency parameter, $c{_0^*} = 0.1$, multiplied by the fraction of non-ionized hydrogen in $H_2$, $ X_{H_2}$, gives the correct normalization of the Kennicutt-Schmidt relation \citep{Christensen2014a}.

We adopt the ``blastwave'' supernova feedback approach \citep{stinson06}, in which mass, thermal energy, and metals are deposited into nearby gas when massive stars evolve into supernovae.  The thermal energy deposited amongst those nearby gas neighbors is 1.5$\times$10$^{51}$ ergs per supernova event.  Subsequently, gas cooling is turned off until the end of the momentum-conserving phase of the supernova blastwave.
The coupling of the supernova thermal energy into the interstellar medium, combined with the turning off of cooling in the affected gas particles, is designed to mimic the effect of energy deposited in the local
ISM by {\it all} processes related to young stars, including UV radiation from massive stars \citep[see][]{Wise2012, Agertz2013}.

It is the recent success of simulations in matching dwarf galaxy properties \citep[e.g.,][]{G10, BZ2014, Shen2014} that allow us to undertake this work. At the dwarf galaxy scale, the different slopes between the observed galaxy stellar mass function and the $\Lambda$CDM predicted halo mass function require that galaxies at halo masses below $10^{11}$ M$_{\odot}$ have gas cooling and star formation efficiencies much lower than those of Milky-Way sized galaxies.  While this trend was historically difficult to produce in cosmological simulations, recent high resolution cosmological simulations that resolve scales on the order of giant molecular clouds can include more realistic models for star formation and feedback, resulting in simulations that can successfully reproduce the observed trends in star formation efficiency \citep{FIRE,Aumer2013,Brook2012,Munshi13,Stinson2013,Simpson2013, G15, Wheeler2015, Christensen2016, Fitts2017}. 
The success of models in reproducing reliable and accurate dwarf galaxies lies in the ability to be able to resolve the impact of baryonic processes on the interstellar medium and star formation \citep{Munshi14, Christensen2014a}.  When this happens, the simulations also simultaneously reproduce additional observed trends in dwarf galaxies, such as cored dark matter density profiles \citep{G12,PG14,diCintio2014a, Maxwell2015, Onorbe2015, Read2016, Dutton2019} and bulgeless disks \citep{brook11, BC2016}. 
 
In the following sections we 
show that the stochasticity of star formation and mass loss of satellites after infall add to the scatter in the  relationship between stellar mass and halo mass. 
We do so with simulations that, {\it a priori}, require no further tuning to successfully match observed properties, including published mass-metallicity relationships \citep{Brooks2007, Christensen2018}, cold gas fractions \citep{Munshi13,Brooks2017},  dark matter profile shapes \citep{G12}, and a host of observed scaling relations for Local Group dwarfs \citep{Applebaum2021}. A future work (Munshi et al., in prep.) will present the Marvel dwarf properties in full detail. 

Individual halos are identified using {\sc Amiga's Halo Finder}\footnote{AHF is available for download at http://popia.ft.uam.es/ AHF/Download.html.} \citep[AHF,][]{gill04,knollmann09}. Throughout this work, the virial radius of a halo is defined to be the radius for which the average halo density is 200 times the critical density of the Universe at a given redshift, 200 $\rho_{crit}(z)$. For subhalos, AHF identifies the virial radius as the point where the lowest density is reached before the density profile increases again due to the contribution from the parent halo. 
For all halos in this work, we trace back the main progenitor to find the peak halo mass that the halo attained, defined as M$_{\rm peak}$.  At each snapshot, the main progenitor is defined to be the halo in the previous step that contains the majority of the particles in the current halo.

\section{Results}\label{Results}

\begin{figure*}
\includegraphics[width=2.0\columnwidth]{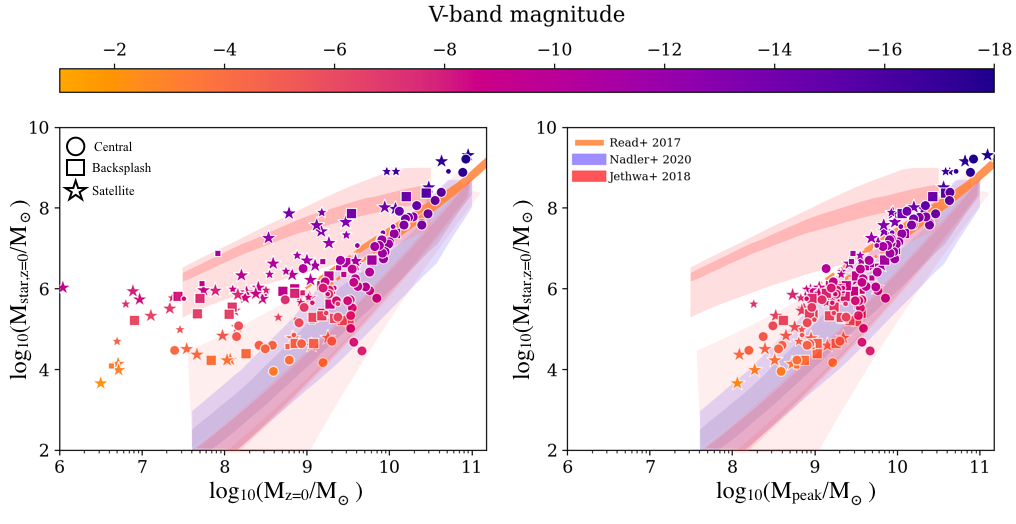}
\caption{{\it Stellar Mass versus Halo Mass.}  Simulated galaxies are color-coded by their $V$-band magnitude at $z=0$.  {\bf Left:} Stellar masses at $z=0$ of the galaxies in the Marvel and Justice League simulations versus their $z=0$ halo mass. {\bf Right:} Stellar masses at $z=0$ versus their peak halo mass. In both panels we display results derived in previous works \citep{Read2017, Jethwa2018, Nadler2020}. The top red contours correspond to $P$(M$_{\rm vir}|$M$_{\rm star})$ and the bottom red contours correspond to $P$(M$_{\rm star}|$M$_{\rm vir})$ from \citet{Jethwa2018}. The stellar masses of classical dwarfs (with M$_V$ brighter than -8) are calculated based on photometric colors \citep[see][]{Munshi13}, while fainter dwarfs use stellar masses directly from the simulations to mimic resolved star counts in UFDs. 
Galaxies represented by circles are isolated galaxies at $z=0$, while galaxies represented by stars are satellites, and squares are backsplash galaxies of the Milky Way-mass halos.  Small points are galaxies from the Near Mint DC Justice League simulations, and larger points are from the Marvel suite and Mint DC Justice League runs. 
While the scatter is decreased by considering M$_{\rm peak}$, the scatter in the relation increases with decreasing halo mass. 
}
\label{fig:smhm}
\end{figure*}

In Figure \ref{fig:smhm}, we show the stellar mass of the simulated galaxies as a function of $z=0$ halo mass (left panel) and as a function of peak halo mass (right panel).  Data points for the simulated galaxies are colored based on the galaxy $V$-band magnitude\footnote{The $V$-band magnitude is calculated in {\it pynbody} \citep{pynbody}, which utilizes the Padova simple stellar population models \citep{Marigo2008, Girardi2010} found at http://stev.oapd.inaf.it/cgi-bin/cmd.} at $z=0$.  The stellar masses for dwarfs brighter than a $V$-band magnitude of $-8$ are calculated using each simulated galaxy's photometric color, as described in \citet{Munshi13}.  For the fainter galaxies, we adopt the simulation stellar mass because the stellar mass of observed UFDs is generally based on resolved star counts rather than photometric color. Circles represent central (isolated) galaxies, squares represent backsplash galaxies, and stars indicate galaxies that are satellites at $z=0$.  Smaller data points are galaxies from the two Near Mint DC Justice League simulations, while the larger data points are from the high resolution Marvel simulations and two Mint DC Justice League simulations.  It can be seen that the brightest galaxies in this sample are roughly LMC-mass \citep[M$_{\rm halo} \sim 10^{11}$ M$_{\odot}$;][]{Kallivayalil2013, Besla2015, Penarrubia2016, Dooley2017}, while the faintest galaxies are UFDs \citep[e.g., M$_{\rm star} < 10^5$ M$_{\odot}$, $V \gtrsim -8$;][]{Simon2019}.

The scatter is greatest if using $z=0$ halo masses (left panel), due to the inclusion of satellite galaxies (shown by stars) that can be stripped of their dark matter after infall to their parent halo.  In some cases subhalos lose more than an order of magnitude in halo mass after infall. 
Stellar masses are more robust, and only those halos with significant halo stripping have lost about a factor of two in stellar mass.  This is consistent with earlier findings that $\sim$90\% of the dark matter mass can be stripped before stars are stripped \citep{Penarrubia2008, Libeskind2011, Munoz2008, Chang2013, BZ2014}.
Considering the peak halo mass (right panel) for the galaxies reduces the scatter, though the lowest mass halos still host almost 2 orders of magnitude in stellar mass. The peak halo mass is typically used in all previous derivations of SMHM relations, including those shown in Figure \ref{fig:smhm}.

The galaxies in Figure \ref{fig:smhm} were all chosen to have resolved star formation histories, which we define to be star formation timescales that span more than 100 Myr.  This choice ensures that, despite multiple supernovae having occurred, the star formation in these galaxies is robust to the feedback.  This choice will exclude galaxies from our simulated sample that undergo a single burst of fast star formation and then quench, but it is not clear if such an event is reliable since it is sensitive to the resolution of the star particles and the feedback prescription.  In practice, our choice of resolved star formation histories leads to a lower limit of 14 star particles in the Marvel and Mint DC Justice League galaxies, and 3 star particles in the Near Mint DC Justice League simulations.  Despite the potentially low number of particles in some Near Mint Justice League galaxies, the SMHM relation of the faintest DC Justice League galaxies blends smoothly into the relation of the higher resolution dwarfs (3 stars is $\sim$1.5$\times$10$^4$ M$_{\odot}$), indicating that there are no obvious resolution effects impacting the SMHM results shown here.  We examine the effect of resolution more carefully in Appendix \ref{Res}.

At the faintest end of the SMHM relation presented in Figure \ref{fig:smhm}, a galaxy cannot be assigned a unique halo mass based solely on its luminosity. As stellar masses decline, the range of halo masses that host a given galaxy increases. Likewise, as halo masses decline, the scatter in the stellar masses of the simulated galaxies increases.

\subsection{Quantifying the Scatter}

In this section we quantify the scatter in our SMHM relation.  We use the M$_{\rm peak}$ results for halo masses, as is commonly adopted for abundance matching or halo occupation studies.  We use the results from these fits to make predictions presented later in the paper.

\subsubsection{Well-resolved Halos}\label{sec:resolved}

Figure \ref{fig:smhm} shows that the scatter in the SMHM relation increases with decreasing halo mass. To quantify the scatter, we assume that stellar masses can be described by log-normal scatter about a mean SMHM relation.
 
At a given halo mass, the scatter in stellar mass is of order 0.3 dex for halos more massive than $\sim10^{10}$ M$_{\odot}$.  The scatter in stellar mass increases to 1.2 dex at the smallest halo masses for these well-resolved halos. With this scatter, there is no longer a single stellar to halo mass relationship for the faintest dwarf galaxies.

\citet{GK2016} demonstrated the impact of scatter in the SMHM relation on the resulting stellar mass function.  They explored both a model in which the scatter is constant as a function of halo mass, and a model in which the scatter increases as halo mass declines.  Clearly, our results favor a model in which the scatter increases toward low halo masses.  \citet{GK2016} quantify the increasing scatter as follows:
\begin{equation}
\sigma = \sigma_0 + \gamma (\rm log_{10} M_{\rm halo} - \rm log_{10} M_1)
\label{eq1}
\end{equation}
where $\sigma$ is the scatter, $\gamma$ is the rate as which the scatter grows, and M$_1$ is a characteristic mass above which the scatter remains constant, at a value $\sigma_0$. We fit our relationship with a broken power law which breaks at M$_1 =10^{10}$ M$_{\odot}$.  Our high mass slope is $1.9$ above the knee
, and the low mass slope is $2.0$ below the knee. Above M$_1$, we assume a constant scatter of $\sigma_0=0.3$ dex, consistent with scatter studies at higher masses \citep{Behroozi2013}, though at the highest masses, the scatter may actually be much smaller \citep{Bradshaw2020}.  Using Eqn. \ref{eq1} with the well-resolved simulated galaxies yields $\gamma = -0.43$.

The knee in our SMHM at M$_1$ corresponds visibly to a knee in our resolved halos (see also Figures \ref{fig:darkfit} and \ref{fig:cosmology}). Our knee corresponds to a halo mass above which all halos in our simulations host a galaxy, but not below.  We discuss the origin of the knee further in Section \ref{sec:reionization}. 
Using only the halos that are well-resolved from a simulation is inherently resolution-dependent.  We address this in the section below. However, the results of using only the well-resolved halos confirm that the scatter is in fact growing, and can be described by log-normal growing scatter. 

\subsubsection{All Halos}\label{sec:all}

Next we extend the discussion to include dark halos: because of limited resolution, some or all of our dark halos may in fact host a galaxy with a stellar mass below our mass resolution. As such, we seek to estimate the underlying SMHM relation in a way that is fully independent of our resolution. To address this, we take the limiting assumption that our unresolved halos would be populated if we were not limited by our mass resolution and model (subgrid) assumptions.  This limiting assumption is supported by earlier work \citep{Jethwa2018, Nadler2020} that shows that all surviving halos down to M$_{\rm peak} = 3 \times 10^8$ M$_{\odot}$, must be occupied in order to match the completeness-corrected abundance of Milky Way satellites.  Assuming fully populated halos steepens the best-fit SMHM relation at low masses (below the knee, where halos start to become unoccupied) due to the inclusion of less massive galaxies at a given halo mass. 

As in the previous section, we model the SMHM as a broken power law. For log(M$_{\rm peak}) \ge 10$, for which all halos in our simulations host galaxies (see Figure \ref{fig:occres}), we fit one slope. For log(M$_{\rm peak}) < 10$, we fit another slope, with growing scatter according to equation \eqref{eq1}. We enforce continuity at the break in both the mean stellar mass, and the scatter, so that we apply equation \eqref{eq1} with $\log\mathrm{M}_1=10$, $\sigma_0=0.3$.

We fit the SMHM relation in a manner that does not assign individual (unresolved) stellar masses to dark halos. Rather, we assume that stellar masses are log-normally distributed along the entire relation, \textit{including for halos that form no stars in our simulations}. ``Dark'' halos are treated as populated galaxies that are merely below our stellar mass resolution. In doing so, we are able to avoid making assumptions about the halo occupation fraction as a function of mass, which in simulations is a function of underlying physics as well as resolution.  We discuss this further below. 

For the SMHM relation below log(M$_{\rm peak}) =10$, we find our parameters $\bm{\theta}$ (i.e., the slope $\alpha$ and scatter growth rate $\gamma$) with the following procedure. 
We divide the SMHM space into $N_k=N_i\times N_j$ bins, with bins $i\in\{1,...,N_i\}$ in $\log(\mathrm{M}_\mathrm{peak})$ and bins $j\in\{1,...,N_j\}$ in $\log(\mathrm{M}_\mathrm{star})$. We count the number of galaxies in each bin $k=(i,j)$. The likelihood of finding the set of simulated galaxies, $\bm{n}$, given our SMHM model parameters, $\bm{\theta}$, is
\begin{equation}
    P(\bm{n}|\bm{\theta})=\prod_{k=1}^{N_k}P(n_{k}|\lambda_{k}),\label{eq:likelihood}
\end{equation}
where $n_{k}$ is the number of simulated galaxies in bin $k$, and $\lambda_{k}$ is the mean number of galaxies in bin $k$ predicted by the model at a given $\bm{\theta}$. The likelihood assumes that $n_{k}$ is complete in the stellar mass range of the bin, with no systematic undercounts due to resolution issues. The counts in each bin can be modeled by a Poisson distribution, so that
\begin{equation}
    P(n_{k}|\lambda_{k})=\frac{\lambda_{k}^{n_{k}}e^{-\lambda_k}}{n_{k}!}.\label{eq:poisson}
\end{equation}

Let $\xi(m|\bm{\Theta})$ give the mean log stellar mass at a given log halo mass, $m$. Since the range of possible stellar masses are normally distributed about $\xi(m)$, the probability that a halo with mass $m$, in bin $i$, will host a galaxy in stellar mass bin $j$ (with bin upper and lower limits of $u$ and $\ell$, respectively), will be
\begin{equation}
    p(j|m,\bm{\theta}) = \int_u^\ell \mathcal{N}\left(\xi(m),\sigma\right)=F(\ell)-F(u),\label{eq:pj}
\end{equation}
where $F(x)$ is the cumulative density function of the normal distribution with mean $\mu$ and standard deviation $\sigma$,
\begin{equation}
    F(x|\mu,\sigma)=\frac12\left[1+\mathrm{erf}\left(\frac{x-\mu}{\sigma\sqrt{2}}\right)\right].\label{eq:erf}
\end{equation}

The expected number of galaxies in bin $k=(i,j)$ according to our model is therefore\footnote{Formally, we would have to calculate $p(j)$ separately for every $m$ in bin $i$. In practice, it is sufficient to calculate it once for each bin, using the mean $m$ of the bin. Our results are insensitive to bin size, indicating this approximation is robust.}
\begin{equation}
    \lambda_k=n_i\,p(j),\label{eq:lambda}
\end{equation}
where $n_i$ is the total number of halos in log(M$_{\rm peak})$ bin $i$, including dark and occupied halos.

\begin{figure}
    \centering
    \includegraphics[width=1\columnwidth]{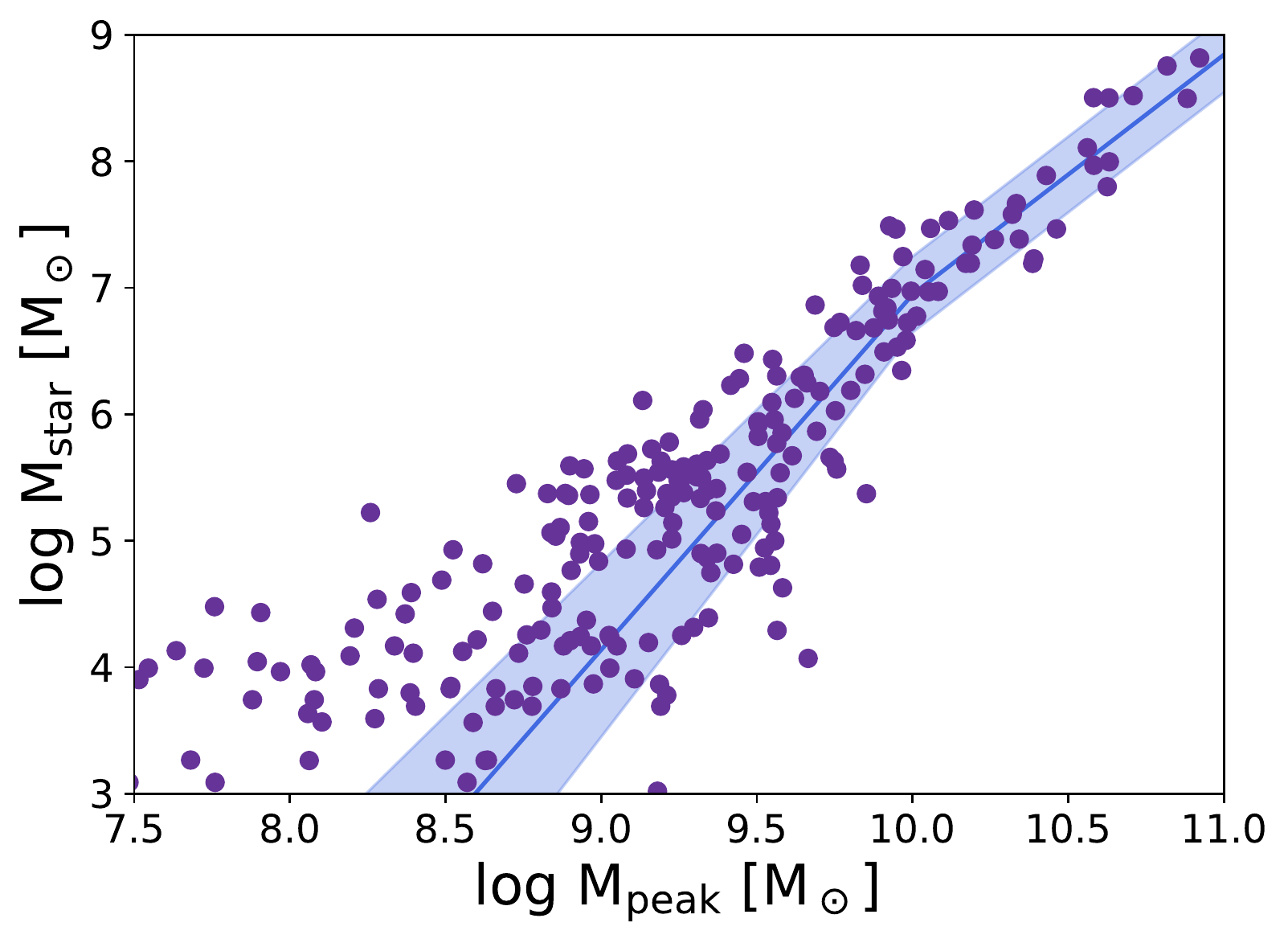}
    \caption{\textit{Stellar mass versus halo mass with fitted relation.} We show the $z=0$ stellar masses and peak halo masses of all simulated galaxies (including galaxies that did not meet the ``well-resolved'' criteria for Figure \ref{fig:smhm}). The solid line shows the best-fit mean relation following the fitting procedure that accounts for the presence of ``dark'' halos; see Section \ref{sec:all} for details. The solid band shows the best-fit log-normal scatter in stellar mass (NB: the band does not represent uncertainty in the best-fit SMHM relation). The best-fit relation is a broken power law separated at log(M$_{\rm peak}) =10$. The high-mass end is described by a slope $\alpha=1.9$ and constant scatter $\sigma=0.3$ dex, and the low-mass end by $\alpha=2.81$ and $\sigma$ given by equation $\eqref{eq1}$ with $\gamma=-0.39$, $\sigma_0=0.3$, and $\log\mathrm{M}_1=10$. Accounting for dark halos steepens the underlying SMHM relation.}
    \label{fig:darkfit}
\end{figure}

To summarize, our likelihood is given by equations \eqref{eq:likelihood} and \eqref{eq:poisson}, with the model galaxy counts given by equation \eqref{eq:lambda}. We fit to galaxy counts\footnote{We found that binning above log(M$_{\rm star})=5$ led to converged results, while including fainter galaxies in the fit led to a steeper slope (we also found varying the lower bounds on log(M$_\mathrm{peak})$ had little effect). We interpret this to mean that we have resolved all galaxies above log(M$_{\rm star})=5$, and below this there is an extended (i.e., non-step function) transition until we no longer resolve any galaxies. While our model accounts for unresolved galaxies, it does so by assuming that the bin counts are complete in our fit region.} in the range $7.5\le$ log(M$_{\rm peak}) \le 11$ and $5\le$ log(M$_{\rm star}) \le9$. We find the best-fitting parameters using the affine-invariant Markov Chain Monte Carlo (MCMC) sampler \textsc{emcee} \citep{Foreman-Mackey2013}, assuming flat priors in the region $0<\alpha<5$ and $-2 < \gamma < 0$. We run $10^4$ steps using 32 walkers, and discard a burn-in period of 500 steps, or ${\sim}15$ times the autocorrelation length.

Using the above procedure, the best-fit SMHM below log(M$_{\rm peak}) =10$ has a slope $\alpha=2.81^{+0.12}_{-0.11}$, with a growing scatter given by equation \eqref{eq1}, with $\gamma=-0.39^{+0.05}_{-0.06}$, $\log \mathrm{M}_1=10$, and $\sigma_0=0.3$ dex (recall, the latter two values were fixed as boundary conditions). The best-fit relation is shown in Figure~\ref{fig:darkfit}.  
As expected, this relation is steeper than just fitting to occupied halos, which bias the underlying relation to higher, resolved masses. Readers may further explore how resolution and scatter can bias the inferred slope of the SMHM in an interactive applet\footnote{Available at \href{https://github.com/emapple/smhm-toy-model}{https://github.com/emapple/smhm-toy-model}.}, which allows users to adjust parameters of a basic SMHM toy model.

\subsubsection{Comparison to Previous Work} 

Our fit to all halos has growing log-normal scatter toward lower halo masses, and reaches $\sigma > 1.0$ dex at log(M$_{\rm peak}) =8$. We are not aware of any other simulations that have allowed scatter to be quantified in the SMHM relation at these low masses, but we can compare our results to simulations of isolated dwarfs at varying masses.  The large scatter in our simulations encompasses the range seen by \citet{Rey2019}, who found that a single ultra-faint dwarf galaxy could vary by an order of magnitude in stellar mass at M$_{\rm peak} = 1.5 \times 10^9$ M$_{\odot}.$  Likewise, the simulated dwarf galaxies of \citet{Revaz2018} and the NIHAO galaxies in \citet{Buck2019} show a comparable amount of scatter in M$_{\rm star}$ at a $z=0$ halo mass of 10$^9$ M$_{\odot}$ as in our Figure~\ref{fig:smhm}. Our galaxies extend to slightly lower stellar masses than \citet{Revaz2018} as might be expected due to our higher stellar mass resolution.  The spread in \citet{Buck2019} is comparable to our Figure~\ref{fig:smhm}, despite their lower resolution, because they plot galaxies with as few as one star particle, while our Figure~\ref{fig:smhm} shows only well-resolved galaxies.

On the other hand, \citet{Fitts2017} simulated 15 field dwarfs with halo masses $\sim10^{10}$ M$_{\odot}$ and a range of halo concentrations, and found a range of M$_{\rm star} = 10^{5-7}$ M$_{\odot}$, as well as one dark halo.  This appears inconsistent with our fixed scatter of 0.3 dex at a similar halo mass.  The FIRE scheme used by \citet{Fitts2017} is stronger and burstier in dwarf galaxies than in the Marvel and Justice League simulations \citep{Iyer2020}, resulting in a steeper SMHM relation in FIRE than presented here, as was also noted by \citet{Revaz2018}.  The distribution of stellar masses in \citet{Fitts2017} is also not well described by a log-normal distribution, but rather with a distribution that peaks at higher M$_{\rm star}$ and has a tail to low M$_{\rm star}$.  This again may be due to the stronger feedback in FIRE, but may also be influenced by their chosen range of halo concentration.

By matching Local Group galaxies to the ELVIS catalog \citep{GK2014b}, \citet{GK2014} found that a large log-normal scatter of 2 dex, constant across M$_{\rm peak}$, alleviated abundance matching discepancies.  If there is no scatter in the SMHM relation, then the number of predicted Local Group and Local Field galaxies in the range $5 < \log($M$_{\rm star}/$M$_{\odot}) < 6$ is larger than currently observed.  Increasing the scatter to 2 dex while steepening the SMHM relation  reduces the predicted number, because the halo mass function is steep in CDM, causing more low mass halos to scatter up than high mass halos to scatter down. Therefore, a given stellar mass is hosted by lower mass halos than might otherwise be expected.  Likewise, \citep{Jethwa2018} find that scatter in the SMHM relation leads to the inference of a shallow SMHM relation if we measure halo mass at a given stellar mass, $P($M$_{\rm vir}|$M$_{\rm star})$.  They find that scatter can help to explain the fact that faint field dwarfs \citep{Ural2015, Read2016b} are sitting in lower mass halos than might otherwise be expected. 

At face value, our results appear to be in tension with \citet{Nadler2020}, who found that the scatter in stellar mass (or luminosity, which is the quantity they used instead) must be small, $\sigma < 0.2$ dex, at all halo masses. There are a couple of differences in our analysis that can bring our results closer together, though not necessarily reconcile the difference.  First, they anchor their growing scatter model at a larger M$_1$ value of 10$^{11}$ M$_{\odot}$, while we anchor at a lower mass of 10$^{10}$ M$_{\odot}$.  If we consider that $\gamma$ is the rate of change of scatter, including the higher mass halos where scatter is essentially constant should decrease $\gamma$ as the scatter does not grow between 10$^{11}$ M$_{\odot}$ to 10$^{10}$ M$_{\odot}$. Fitting all of our resolved halos with a single slope and growing scatter model anchored at 10$^{11}$ M$_{\odot}$ (instead of growing scatter only below the knee, anchored at 10$^{10}$ M$_{\odot}$), our value for $\gamma$ decreases from -0.39 to to -0.21.  

Also, \citet{Nadler2020} examine scatter in luminosity as a function of v$_{\rm peak}$, while we are tracing based on M$_{\rm peak}$.  Previous works have shown that M$_{\rm star}$ correlates more strongly with v$_{\rm peak}$, due to the effects of halo assembly bias \citep[e.g.,][]{Chaves2016, Reddick2013}.  This correlation should lead to smaller scatter in the  M$_{\rm star}$ - v$_{\rm peak}$ relation than in the SMHM relation.  We have verified that our scatter is reduced when using v$_{\rm peak}$, e.g., decreasing our scatter above the knee from 0.3 dex to 0.17 dex.  However, our scatter still increases below the knee, blowing up to over 1 dex at the lowest v$_{\rm peak}$ values we trace, as with M$_{\rm peak}$.  Thus, while the change in variables can explain some difference in the results, it does not explain it all.  

Finally, \citet{Nadler2020} found that the scatter must be small to reproduce the observed LFs of Pan-STARRS1 and the Dark Energy Survey \citep[DES,][]{DES1, DES2}; with large scatter, low-mass halos host satellite galaxies that scatter to observable luminosities too often, and they were unable to reproduce the satellite LF of the Milky Way.  The Milky Way may \citep{Carlsten2020, Mao2020} or may not \citep{Wang2021} have a typical satellite LF for a galaxy of its luminosity, but the four Justice League simulations used here have been shown to match the range of satellite LFs of observed Milky Way-mass galaxies \citep{Akins2020, Applebaum2021}.  \citet{Nadler2020} derived their constraints using only two Milky Way realizations, which could impact their scatter results.  A full accounting of this discrepancy requires further exploration.

\subsection{Scatter and the Stellar Mass Function}\label{popmassfunc}

\begin{figure*}
\centering
\includegraphics[width=1.5\columnwidth]{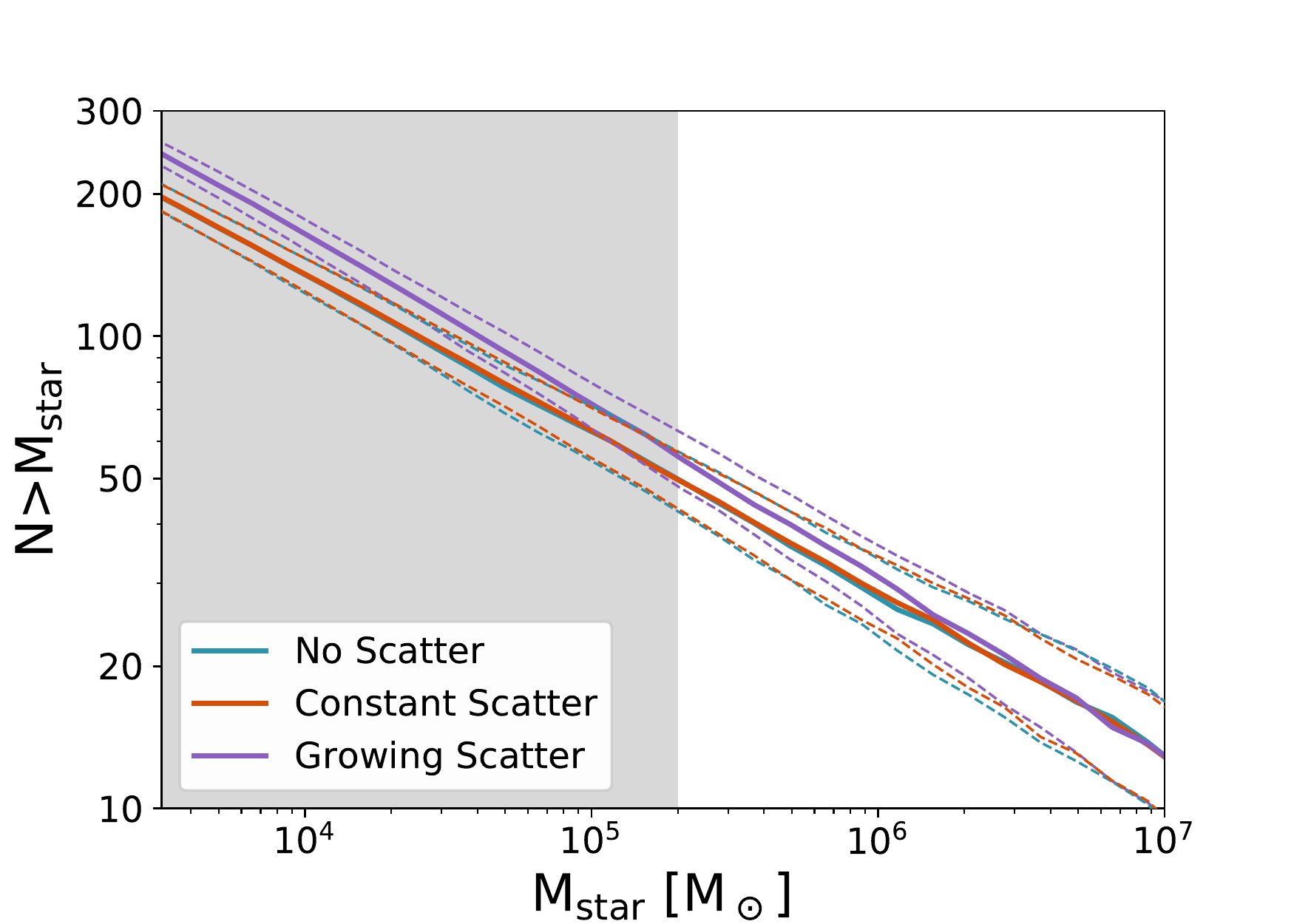} 
\caption{{\it The effect of SMHM scatter on the predicted stellar mass function.} Constant Scatter is very similar to No Scatter, both in slope and normalization.  Growing Scatter, however, steepens the SMF.   We generate each SMF 1000 times, and show in the solid lines the median relation and in the dashed lines the inner 68\% range. We have additionally scaled each SMF so that they have 12 galaxies above $10^7$~M$_\odot$, consistent with the number within 1~Mpc of the Milky Way. 
The shaded region represents the approximate discovery space of UFD galaxies. }
\label{fig:elaad}
\end{figure*}

In this subsection, we use our predicted slopes and scatter in the SMHM relation in order to calculate a predicted stellar mass function (SMF). To do this, we draw 1000 random halos with $10^{7.8}\leq\mathrm{M}_\mathrm{\rm peak}\leq10^{11.5}$ from the ELVIS catalogs \citep{GK2014}, and populate them with stellar masses according to our predicted SMHM slopes and scatter.  The resulting SMFs are shown in Figure \ref{fig:elaad}  
for three cases: (i) no scatter in the SMHM relation (``No Scatter''), (ii) constant log-normal scatter of 0.6 dex in M$_{\rm star}$ at a given M$_{\rm peak}$ (``Constant Scatter''), and (iii) adopting our results for growing scatter from Section \ref{sec:all} (``Growing Scatter'').  For the Growing Scatter case, recall that we adopt a slope of the SMHM $\alpha = 1.9$ for halos from 10 $<$ log(M$_{\rm peak}) < 11.5$.  We adopt a steeper slope of $\alpha = 2.8$ at lower halo masses, utilizing the slope derived from probabilistically populating halos below our resolution limit.  The scatter is constant at 0.3 dex above M$_1 = 10^{10}$ M$_{\odot}$, and linearly grows below this halo mass, i.e., $\gamma = -0.38$. We generate each SMF 1000 times, and show in the solid lines the median relation with the inner 68\% range indicated by the dashed lines. We have scaled each SMF so that they have 12 galaxies above $10^7$~M$_\odot$, consistent with the number within 1~Mpc of the Milky Way.

\citet{GK2016} demonstrated that large scatter in the SMHM relation impacts the observed SMF.  Due to the rapidly rising halo mass function, there are more small halos available to scatter to larger stellar masses than large halos to scatter to lower stellar masses, an effect that is increasingly noticeable as scatter increases.  
The change from the No Scatter case to the Constant Scatter case amounts to a uniform shift upwards in the number of galaxies of a given stellar mass in log space in the SMF.  This shift upwards disappears after we rescale the SMF to $N = 12$ for M$_\mathrm{\rm star}$ $> 10^7$ M$_\odot$, which is why the No Scatter and Constant Scatter mean SMF lie on top of each other in Figure \ref{fig:elaad} (note that this will occur no matter the magnitude of the scatter chosen for the Constant Scatter case). If scatter is not constant, however, and grows with decreasing halo mass, then the shift in the SMF is not uniform.  Hence, the Growing Scatter SMF steepens in Figure \ref{fig:elaad}. 

Our scatter grows in such a way that it only reaches relatively large scatter ($> 1$ dex) at very low halo masses that host primarily UFD galaxies. %
This is where the scatter begins to have a noticeable impact on the stellar mass function, compared to the Constant Scatter and No Scatter cases.  
The mass range where there is greatest difference between SMFs corresponds to luminosities where current observations are incomplete, but are being or will be probed by surveys like DES, HSC-SSP, or with the Vera Rubin Observatory\citep{tollerud08, Walsh2009}.  In Figure \ref{fig:elaad}, we shade the mass region of the new dwarf satellites that have been found in the first two years of the DES \citep{DES1, DES2}. \citet{tollerud08} estimates that such faint dwarf galaxies should be observed out to $\sim$1 Mpc by the Vera Rubin Observatory's LSST after the full co-added data are collected.  Thus, the slope of the SMF, when complete, can possibly constrain the magnitude of scatter in the SMHM, and possibly the smallest halo which hosts a galaxy.

\section{Discussion}\label{Discuss}

In this section we discuss various factors that might influence our predicted SMHM relation.

\subsection{Occupation Fraction}\label{focc}

The occupation fraction in the real Universe--- i.e., the fraction of halos at a given mass that host a galaxy---is dependent on many physical processes (discussed further below), including the physics of star formation, gas cooling, self-shielding, and the strength and timing of reionization. Quantifying this ``intrinsic'' occupation fraction in a cosmological simulation would require the ability to resolve large numbers of both the smallest halos that can host galaxies as well as the smallest stellar mass that can constitute a galaxy, which is still observationally unconstrained. Given that galaxies have been observed with masses as low as ${\sim}10^2$~M$_\odot$ \citep[e.g.][]{DES2, Homma2018, Longeard2018}, cosmological simulations to date do not have the ability to reliably resolve the intrinsic occupation fraction. Rather, whether a halo is ``occupied'' or ``dark'' in the simulations is subject to resolution.

\begin{figure}
\includegraphics[width=1.0\columnwidth]{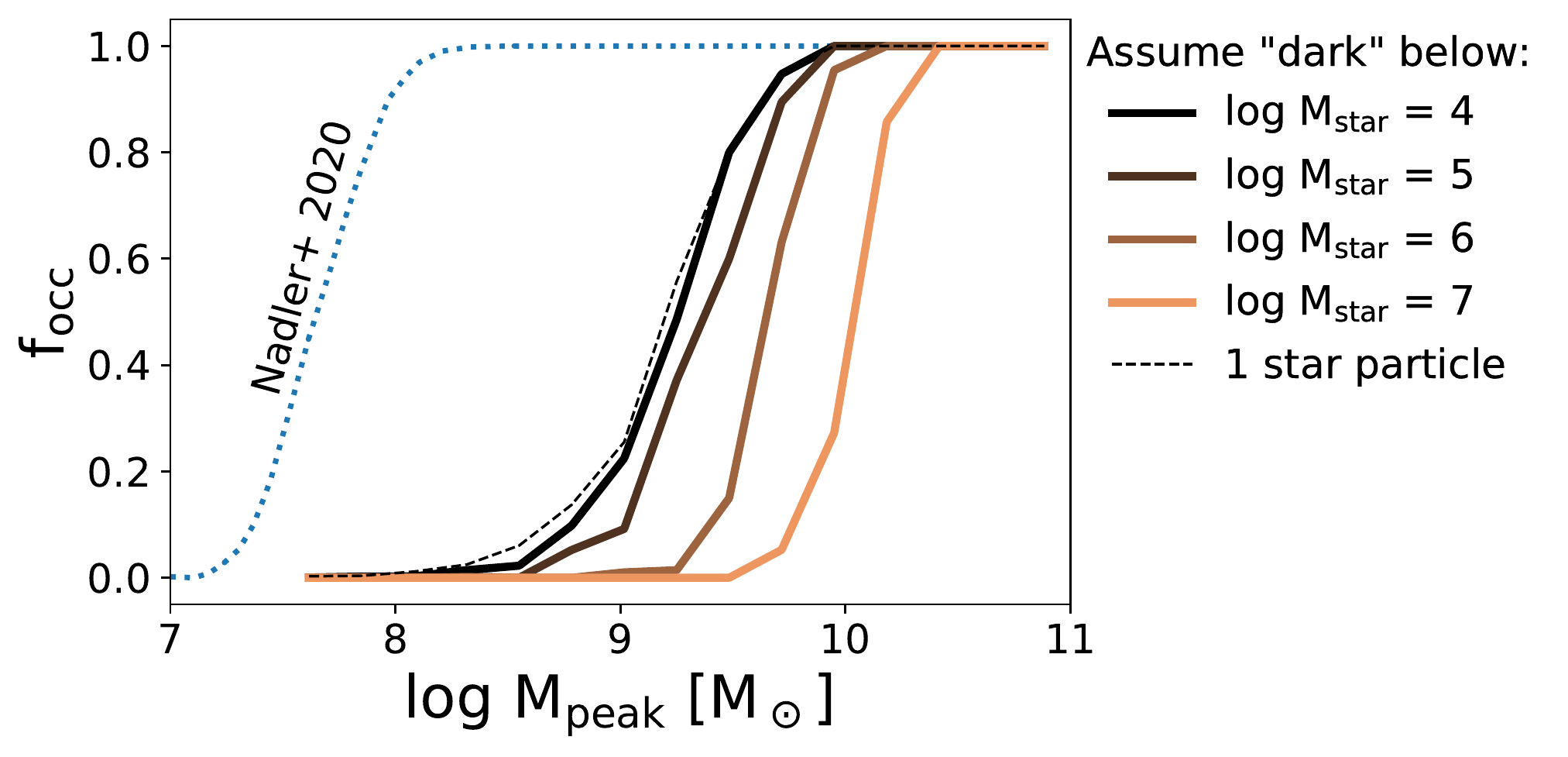}
\caption{{\it The dependence of occupation fraction on simulation resolution.} As simulations have finite mass resolution, the minimum mass of an occupied halo is inherently tied to resolution. We demonstrate that the simulated occupation fraction changes significantly when the minimum resolvable stellar mass (and in effect, the mass resolution) is varied. Both the peak halo mass where galaxies go ``dark'' and how sharply the curve declines to zero are affected. The sharpness of the decline increases as resolution decreases as a result of the decreasing scatter in the SMHM relation at higher masses. In the dotted line we show the occupation fraction that we would infer from the simulations if we assumed that halos that host at least 1 star particle contain galaxies. For comparison, we show the occupation fraction inferred by \citet{Nadler2020}, which is constrained by Milky Way observations.}
\label{fig:occres}
\end{figure}

We show this simplistically in Figure \ref{fig:occres} by choosing various threshold stellar masses below which all dark matter halos are devoid of stars (i.e., don't host a galaxy). This, in effect, is varying the mass resolution for galaxies in our simulations.  We also indicate (black dashed line) the occupation fraction we would infer from our simulations if we did not recognize that it is resolution dependent, assuming all halos with at least one star particle host a galaxy.\footnote{We use all halos with one star particle despite the two different resolutions used in this work.  The lower resolutions simulations will reach zero occupation at a slightly higher M$_{\rm peak}$ than the higher resolution simulations, but the number of halos in that mass range is so large that it has little impact on $f_{\rm occ}$.}  As we increase the threshold stellar mass, not only does the halo mass where galaxies go ``dark'' increase, the sharpness of the simulated occupation fraction becomes closer to a step function.  The sharper decline comes from the smaller scatter in the SMHM relation at higher stellar masses.%

Both \citet{Jethwa2018} and \citet{Nadler2020}, using observations combined with a model for the galaxy-halo connection, find that essentially all halos with peak mass above ${\sim}3 \times 10^8$~M$_\odot$ need to host a galaxy in order to be consistent with the completeness-corrected observations of dwarf galaxies. 
Jethwa et al.~followed galaxies down to $M_V = -1.5$, while Nadler et al.~modeled galaxies as faint as $M_V = 0$ (M$_\mathrm{\rm star}\sim100$~M$_\odot$). Our simulations are unable to resolve galaxies this faint. The inferred occupation fraction derived by \citet{Nadler2020} is shown by the blue dotted line in Figure \ref{fig:occres}. While our higher resolution simulations have $z=0$ star particle masses as low as 174 M$_{\odot}$, the occupation fraction inferred from the simulations drops to zero at substantially higher M$_{\rm peak}$ than Nadler et al.  This is because we start to be unable to resolve galaxies that have low stellar mass that would reside in halos with M$_{\rm peak} < 10^9$ M$_{\odot}$.  This motivates the approach we have taken in this work, in which we do not impose an occupation fraction {\it a priori}, nor do we impose the occupation fraction explicitly defined by our simulations, as we know this to be resolution dependent. Instead, we assume that stellar masses are log-normally distributed along the entire SMHM relationship, as our resolved halos show a log-normal growing scatter. In doing so, we treat dark halos as populated, but populated below our resolution limit.
This approach then allows us to make predictions for the SMF to highlight the effect of scatter in the SMHM relation, removing the effect of a resolution-dependent occupation fraction.  

We note that this approach of treating the occupation fraction as resolution dependent is very different to what has been assumed in the past.  It has long been thought that reionization should remove baryons from low mass halos, leaving a population of low mass halos that are completely ``dark'' and without a galaxy.  Early works suggested that the halo mass cutoff for galaxy formation was fairly massive \citep[e.g.,][]{benson02, somerville04}.  The idea that reionization should imprint itself at a halo mass that is somewhere on the order of 10$^8$ to 10$^{10}$ M$_{\odot}$ has persisted, leaving simulators \citep[e.g.,][]{Sawala2015} to treat their halo occupation fractions as predictions rather than recognizing the resolution dependence inherent in the results.  It is only the recent reconciliation of observed UFD galaxy counts with halo counts that have forced a re-examination of this assumption \citep[see also][]{Kim2018, Newton2018, Read2019}.

However, if some halos with peak mass greater than 10$^{8}$ M$_{\odot}$ are truly dark, then this will have a strong impact on the resulting SMF.  For example, \citet{Munshi2019} showed that the star formation prescription we use in this work, in which H$_2$ must be present for stars to form, inhibits the formation of UFD galaxies in halos with log M$_{\rm halo} > 8.5$.  They also showed that a different star formation prescription, which required high densities and low temperatures but not necessarily the presence of H$_2$, allowed galaxies to form in halos at least an order of magnitude lower in mass.  It is not clear that either of these models is truly reflective of star formation in UFDs, but the result demonstrates that there may be an influence beyond reonization that impacts the halo occupation distribution.  A comparison of our Figure \ref{fig:elaad} with Figure 6 of \citet{Munshi2019} shows that occupation fraction can have a much stronger impact on the resulting SMF than the large scatter in the low mass SMHM relation we have found in this work.

\subsection{Reionization}\label{sec:reionization}

Our SMHM relation has a bend at $\sim 10^{10}$ M$_{\odot}$ below which the slope of the SMHM steepens.  This is true independent of cosmology or resolution, as can be seen in Appendix \ref{Res}, where the knee is readily apparent. This knee is also reflected in the full sample in Figure \ref{fig:smhm}, though the large number of dwarf galaxies and large scatter at low masses make it less obvious visually.  This knee in the relationship, dividing the SMHM into two slopes, is an expected result of reionization. There will be some halo mass (the knee) below which star formation is suppressed, either due to gas loss, gas suppression, and/or gas ionization.  \citet{Gnedin2012} finds that the contribution of ionized gas becomes significant in galaxies with $v_{\rm max} \sim 40-50$ km s$^{-1}$.  Our knee at 10$^{10}$ M$_{\odot}$ corresponds to $v_{\rm max} \sim 40$ km s$^{-1}$.  The onset of reionization more strongly impacts the earlier forming (i.e., lower mass) halos \citep{Bose2018, Bose2020, BL2020}.  The impact of reionization at a given halo mass may depend on local density and timing of the onset of reionization (discussed below), but it follows that there should be two different slopes in the SMHM relation, with a knee  below which reionization is suppressing star formation in halos (see also \citealt{GK2019}).

As discussed in \citet{Munshi2019}, many of the low mass halos in our H$_2$ star formation model are not able to form stars before reionization prevents them from doing so.  This implies that our results are sensitive to our selected reionization model.  
We have adopted the same model in all simulation volumes, following \citet{HM2012}.  However, \citet{HM2012} has been shown to heat the IGM earlier ($z \sim 15$) than it should \citep{Onorbe2015}, potentially making the impact of reionization particularly strong on our results.  

Arguably the more important limitation of our reionization model is the fact that it is uniform throughout the simulation volume.  This is common for cosmological galaxy simulations, as the radiative transfer required to explicitly follow patchy reionization is computationally expensive.  Simulations that self-consistently model reionization find that the baryon fraction of low mass halos is highly dependent on the timing of reionization due to local densities, with suppression due to reionization occurring earlier in higher density environments \citep{Wu2019, Katz2020, Ocvirk2020}. The Marvelous Dwarfs are $\sim1.5-7$ Mpc away from a Milky Way-mass galaxy, meaning that, in a more realistic reionization scenario, they are in a lower density region that may not ionize as early as the higher density regions surrounding the Justice League Milky Ways.  The fact that our full sample blurs out the knee in the SMHM relation more than near individual massive galaxies (compare Figure \ref{fig:smhm} to Appendix \ref{Res}) likely points to environmental dependence on reionization at a fixed halo mass even with our spatially uniform UV background.  However, a more realistic reionization model may allow some of our dark halos in our lower density regions to form stars.   \citet{BL2020} explicitly look at the effect of reionization on the $z=0$ occupation fraction, and find that it varies with how early or late reionization begins. 
Understanding the impact of reionization will require further simulations and study, but will be essential to explore the uncertainties that will impact the interpretation of LSST observations. 

\subsection{Star Formation and Feedback Model}

As discussed in Section \ref{focc}, \citet{Munshi2019} finds that there is a significant reduction in overall efficiency of star formation in simulated UFD galaxies when adopting a non-equilibrium H$_2$-based star formation prescription relative to a prescription that adopts a temperature-density threshold. The reduction in star formation in the H$_2$ model is due the long formation times of H$_2$ at low metallicities in low mass halos.  This yields a significant difference in the number of predicted galaxies at low masses and results in a different predicted SMF between the two star formation recipes. In this paper, we adopt the same H$_2$ based star formation model as in \citet{Munshi2019} across all 8 volumes, and use our large sample of dwarf galaxies to test the effect of the slope and scatter of the SMHM on the predicted SMF.  Figure \ref{fig:elaad} shows that there is no appreciable change in the faint-end slope of the SMF or number of predicted UFDs when considering No Scatter versus the Constant Scatter case.  However, in the Growing Scatter case, the SMF is steepened towards lower masses.  A comparison with Figure 6 of \citet{Munshi2019} shows that the occupation fraction imposed by star formation prescription can have a much larger effect on the number of predicted UFDs: the large scatter in the SMHM relation derived in this paper impacts the number of UFDs by $\sim$25\%.  The star formation prescription contributes a comparable change {\it if} the slope of the SMHM relation is unchanged.  If the star formation prescription also changes the slope of the SMHM relation, then the number of UFDs can be a factor of 2-3 different.

Alternatively, \citet{Byrne2019} demonstrate that tying star formation to shielded gas rather than the presence of H$_2$ allows stars to form in lower density gas.  This may change the resulting SMHM in the regime where dwarf galaxies do not self-regulate.  
\citet{Latif2019} find H$_2$ self-shielding is critical: in addition to the delay in star formation, the collapse and evolution of halos is tied to the strength of the UV background.  Furthermore, our H$_2$ model does not let stars form (at our current resolution) in halos below $10^8$ M$_{\odot}$ at $z=6$.  \citet{Skinner2020} and \citet{Latif2019} show halos below this mass should be forming Pop III stars, which are not included in our model.  Both emphasize the balance between UV flux and self-shielding that sets the halo mass that can form stars, and thus the $z=0$ occupation fraction. These results emphasize that the number of newly discovered UFDs by LSST will place constraints on both the process of star formation and the UV background at high redshift.

In addition to star formation recipe, feedback strength and implementation varies between simulations. Galaxy stellar masses are sensitive to specific feedback implementations. For example, \citet{Agertz2020} find that varying feedback models changes the stellar mass of their test halo by over 1 dex, but still within the scatter of our SMHM relation. In simulations run with \textsc{ChaNGa} and \textsc{Gasoline}, superbubble feedback \citep{Keller2014} leads to a factor of ${\sim}2$ reduction in stellar mass in Milky Way-like halos \citep{Keller2015}. \citet{Mina2020} simulate dwarf galaxies using superbubble feedback, and they find that it has a varying effect on the SMHM relation. A larger sample is needed to assess any potential systematic effect of superbubble feedback on dwarf stellar masses. Specific results may also be sensitive to feedback details like the inclusion of radiative feedback: For example, \citet{Smith2020a} shows that ionizing radiation reduces supernova clustering, leading to suppressed supernova-driven outflows, while \citet{Smith2019} stress that the strength of supernova feedback in dwarfs may also be highly dependent on ISM turbulence, runaway massive stars, and early stellar feedback.  To constrain these varying physical processes, it may be that a more detailed accounting of the observed stellar mass fraction as a function of halo mass \citep[e.g.,][]{Read2019, Romeo2020} is required.

Finally, coupled with non-supernova feedback, the effects of IMF sampling on star formation are more difficult to predict \citep{Smith2020b}. \citet{Applebaum2020a} shows that implementing a stochastically populated IMF in \textsc{ChaNGa} can  systematically lower stellar masses in UFDs. 

\bigskip
In summary, how galaxies populate dark matter halos in simulations is not only resolution dependent, but also tied to both star formation model and reionization.  Since our choice of star formation model and our reionization model can affect the slope and scatter of halos we predict, we are unable to say that the number of UFDs predicted in this paper is exactly what LSST should expect to find. However we {\it can} constrain the effect of scatter in the SMHM relation on predictions compared to a no scatter case as long as our underlying models and assumptions remain constant.

\section{Summary}\label{Summary}
In this paper, we use a large sample of extremely high resolution simulated dwarf galaxies (LMC-mass to ultra-faint dwarfs) in a range of environments in order to predict the SMHM relation at low halo masses. This is the first prediction of the SMHM relation for dwarf galaxies this low in mass at $z=0$, and the first time that the scatter at the low mass end has been robustly quantified. In doing so, we demonstrate that (1) derived SMHM relations cannot be simply extrapolated from higher masses to lower masses, and thus (2) the halo mass of a faint dwarf galaxy cannot be inferred {\it a priori}. 

Our SMHM relation is best fit by including a break in the slope of the SMHM relation.  This break corresponds to a visible knee in the relation, above which all halos are occupied by simulated galaxies, and below which we start encountering dark halos.  This knee is the natural consequence of reionization.  The details of the reionization model will matter for the exact location of the knee.  In this work, in which we adopt a uniform UV background based on \citet{HM2012}, we find a break at a halo mass of 10$^{10}$ M$_{\odot}$.  The slope above the knee is $\alpha = 2.0$, and below the knee it steepens to $\alpha = 2.8$.

The scatter in stellar mass at a given (peak) halo mass is constant above the knee at 0.3 dex.
Below the knee, some halos may truly be dark due to reionization, but it is also likely that these low mass halos can host galaxies that are below our mass resolution limit.  
Assuming that all halos are occupied, and that the log-normal scatter that grows linearly about our best fit slope of $\alpha = 2.8$ below the knee, the rate of increase in scatter is quantified by $\gamma$, where $\gamma = -0.38$ (see Eqn.~\ref{eq1}).

When using $z=0$ halo masses, the scatter in the relation is much larger, due to the inclusion of satellite and backsplash galaxies in the SMHM relation. The satellites of both our dwarf and Milky Way galaxies as well as backsplash galaxies can lose substantial mass after infall, including satellites that lose an order of magnitude or more in halo mass.  The results presented above instead use peak halo mass, similar to earlier abundance matching studies, which significantly tightens the SMHM relation.

Our best fit SMHM relation slopes and scatters can be used to stochastically populate theoretical models at low masses instead of relying on abundance matching.  We demonstrate this by populating a halo mass function and showing the predicted SMF (see Figure \ref{fig:elaad}), and exploring the effect of scatter on the SMF.  The resulting SMF is essentially unchanged when considering the Constant Scatter and No Scatter cases, due to the fact that the SMF is normalized to have 12 galaxies above M$_{\rm star} = 10^7$ M$_{\odot}$ within 1 Mpc, comparable to the Local Group. On the other hand, the faint-end slope of the SMF is steepened in the case of Growing Scatter. More UFD galaxies are predicted if there is Growing Scatter in the SMHM relation.  The luminosity/mass range where this can be tested is currently being probed by surveys like DES, which discovered 16 new UFDs in its first two years \citep{DES1, DES2}, and the HSC-SSP, which has discovered three dwarfs \citep{Homma2016, Homma2018, Homma2019}.  Additional dwarfs should be discovered by the Rubin Observatory's LSST when it comes online, allowing the SMF in this range to be probed observationally.

This work is part of a series that has explored uncertainties in modeling UFD galaxies.  \citet{Applebaum2020a} showed that adopting a stochastic IMF can alter the stellar masses of UFD galaxies.  \citet{Munshi2019} showed that two commonly adopted simulation star formation prescriptions can yield different low mass SMHM relations and occupation fractions.  In this work we have discussed the role of occupation fraction, reionization model, and star formation model on our resulting SMHM relation.  Because of these uncertainties, we are unable to say that our predicted SMF is what LSST is going to find.  However, by systematically exploring these effects we can quantify the uncertainties in the predicted SMF given known unknowns, and begin to search for ways to break degeneracies.

Despite the caveats, the large intrinsic scatter in our simulated SMHM demonstrates that the commonly-adopted assumption of a monotonic relationship between stellar mass and halo mass that is adopted in abundance matching breaks down at low masses.  Reionization should impose a break in the relation at low masses.  At the faintest end of the SMHM probed by our simulations, a galaxy cannot be assigned a unique halo mass based solely on its stellar mass or luminosity. The lack of a monotonic relation between stellar mass and halo mass has implications for interpreting observations of dwarf galaxies.

\section*{Acknowledgements}
FDM acknowledges support from the Vanderbilt Initiative for Data-Intensive Astrophysics (VIDA) through a  VIDA Postdoctoral Fellowship, and from the University of Oklahoma. FDM also acknowledges that Oklahoma's land is the traditional territory of the Osage, Cherokee, Choctaw, Chikasaw and Patowatomi Peoples who have lived here since time immemorial. FDM and AMB acknowledge support from HST AR-13925 provided by NASA through a grant from the Space Telescope Science Institute, which is operated by the Association of Universities for Research in Astronomy, Incorporated, under NASA contract NAS5-26555. FDM and JPS acknowledge support from NSF grant PHY-2013909. AMB acknowledges support from HST AR-14281.  EA and AMB acknowledge support from NSF grant AST-1813871. EA acknowledges support from the National Science Foundation (NSF) Blue Waters Graduate Fellowship. CRC acknowledges support from the NSF under CAREER grant No. AST-1848107.  Resources supporting this work were provided by the NASA High-End Computing (HEC) Program through the NASA Advanced Supercomputing (NAS) Division at Ames Research Center.  This research also made use of the NSF supported Frontera project operated by the Texas Advanced Computing Center (TACC) at The University of Texas at Austin. This research was supported in part by the National Science Foundation under Grant No.~NSF PHY-1748958.

The authors thank Yao-Yuan Mao, Ethan Nadler, Martin Rey and Priya Natarajan for assistance and helpful comments which improved this manuscript. 

\appendix
\section{The effect of Cosmology and Resolution on Scatter}\label{Res}
\begin{figure}[!tbp]
  \centering
  \begin{minipage}[b]{0.7\textwidth}
    \includegraphics[width=\textwidth]{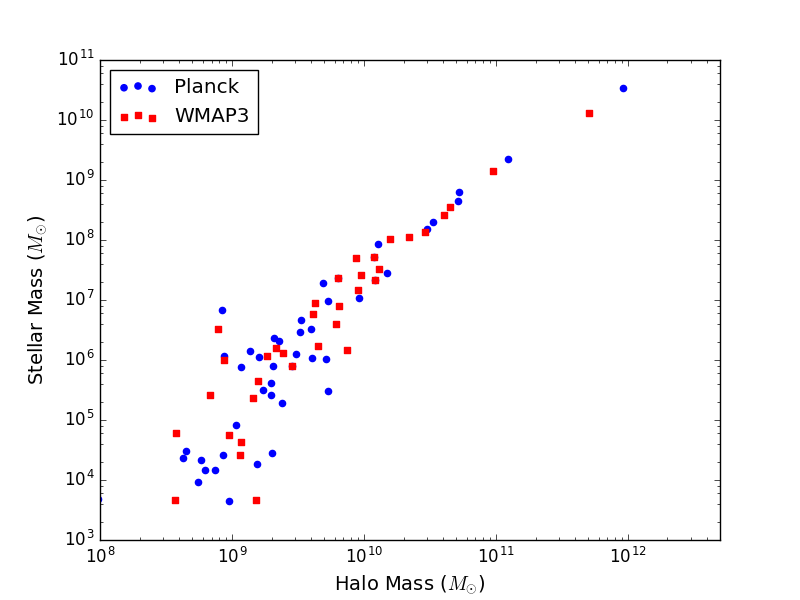}
  \end{minipage}
  \hfill
  \begin{minipage}[b]{0.7\textwidth}
    \includegraphics[width=\textwidth]{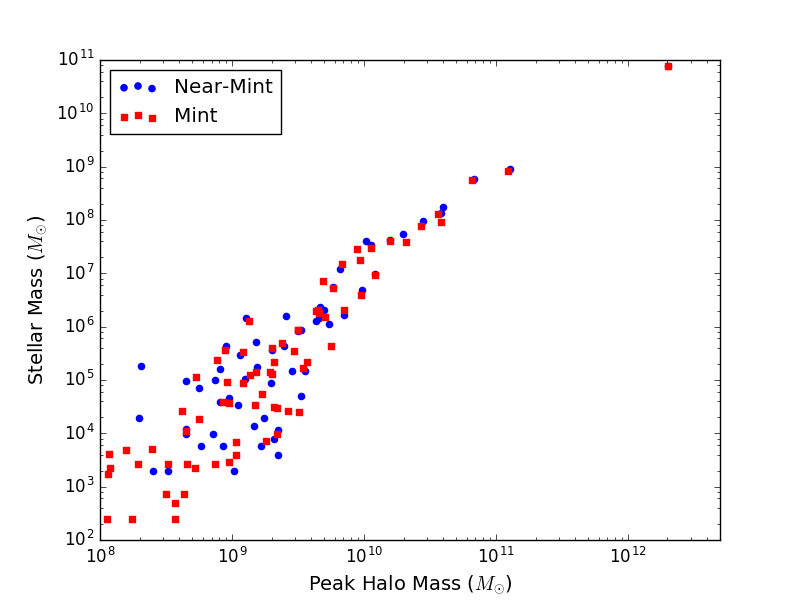}
  \end{minipage}
  \caption{{\it SMHM relationship for various versions of Sandra.} Top Panel: We compare $z=1.4$ results from both the WMAP3 cosmology and the Planck cosmology at NM resolution to show convergence between different cosmologies.  Bottom Panel: We show $z=0$ results using a Planck cosmology between both the NM run and Mint run. Both resolutions converge to a similar SMHM. Both panels use simulation stellar masses, rather than photometric.}
\label{fig:cosmology}
\end{figure}


As mentioned in Section \ref{Sims}, the Marvel dwarfs are run within a WMAP Year 3 cosmology \citep{spergel07} while the Justice League simulations are run with a Planck cosmology \citep{Planck2016}. In the top panel of Figure \ref{fig:cosmology} we examine the effect of different cosmologies on the resulting SMHM relation.  The top panel of Figure \ref{fig:cosmology} compares the $z=1.4$ results for Sandra's (one of the Justice League Milky Ways) SMHM in both WMAP3 cosmology (red points) and in Planck cosmology (blue points), at the same NM resolution.  We find that the SMHM relation is consistent across cosmologies. To quantify, {\it we fit a single slope and scatter across the entire range of masses.} For reference, if we fit a single slope and scatter for our full sample instead of the two component fit as in Section \ref{sec:all}, the slope is $\alpha = 2.0$ with a scatter of $\gamma = -0.21$. The slope and scatter of the $z=1.4$ SMHM relations with WMAP3 and Planck cosmomolgies are consistent with each other and with the well-resolved sample presented in Figure \ref{fig:smhm} to within one sigma ($\alpha = 1.97$/$\gamma = -0.2$ and $\alpha = 2.0$/$\gamma = -0.2$, respectively). 

In the bottom panel of Figure \ref{fig:cosmology} we compare the $z=0$ NM and Mint Sandra simulations in a Planck cosmology to test the effect of resolution. The stellar masses converge between the two simulations \citep[see also][]{Applebaum2021}, although with the increased Mint resolution we resolve more UFDs. Here too we find similar slopes and scatters between the two resolutions, within one sigma of each other ($\alpha = 2.02$/$\gamma = -0.2$ and $\alpha = 2.05$/$\gamma = -0.2$ for NM and Mint, respectively). Since the cosmological comparison is at a higher redshift and both comparisons (resolution and cosmology) only include one simulation, we compare slope and scatter in the above substantially simplified form.

\bibliographystyle{mn2e}
\bibliography{bibref}

\end{document}